\documentclass{nature}

\usepackage{graphicx}

\usepackage{times}
\usepackage{amsmath}
\usepackage{upgreek}

\usepackage{xcolor}

\usepackage{soul}
\setstcolor{red}

\title{On-chip quantum sensing of Kondo spins in a high-mobility quasi-one-dimensional nanoconstriction}

\author{Shun-Tsung Lo$^{1,2\ast}$, Che-Cheng Wang$^{2}$, Sheng-Chin Ho$^{2}$, Jun-Hao Chang$^{2}$, Ming-Wei Chen$^{2}$, G. L. Creeth$^{3}$, L. W. Smith$^{2}$, Shih-Hsiang Chao$^{2}$, Yu-Chiang Hsieh$^{2}$, Pei-Tzu Wu$^{1}$, Yi-Cheng Wu$^{1}$, Chi-Te Liang$^{4}$, M. Pepper$^{3}$, J. P. Griffiths$^{5}$, I. Farrer$^{5,6}$, G. A. C. Jones$^{5}$, D. A. Ritchie$^{5}$, \& Tse-Ming Chen$^{2,7\ast}$}

\begin{document}

\maketitle

\begin{affiliations}
 \item Department of Electrophysics and Center for Emergent Functional Matter Science, National Yang Ming Chiao Tung University, Hsinchu 300, Taiwan 
 \item Department of Physics, National Cheng Kung University, Tainan 701, Taiwan
 \item Department of Electronic and Electrical Engineering, University College London, London WC1E 7JE, United Kingdom
 \item Department of Physics, National Taiwan University, Taipei 106, Taiwan
 \item Cavendish Laboratory, J J Thomson Avenue, Cambridge CB3 0HE, United Kingdom 
 \item Department of Electronic and Electrical Engineering, University of Sheffield, Mappin Street, Sheffield S1 3JD, United Kingdom
 \item Center for Quantum Frontiers of Research \& Technology (QFort), National Cheng Kung University, Tainan 701, Taiwan
\end{affiliations}

\noindent $^\ast$To whom correspondence should be addressed;\\
E-mail: stlo@nycu.edu.tw; tmchen@phys.ncku.edu.tw.\\

\textbf{The precise nature of Kondo spins has remained enigmatic when extended to multiple spin impurities or, more intriguingly, when the localized spin itself may already be the consequence of many-body interactions in a presumably-delocalized open nanoconstriction, such as a quantum point contact (QPC). It is experimentally challenging to distinguish the Kondo state from other coexisting many-body spin states in such a strongly correlated system. Here we lithographically define an all-on-chip electronic resonator (ER) and a QPC in a high-mobility GaAs/AlGaAs heterostructure transistor. Local Kondo screening of the QPC spin and nonlocal spin singlet across the ER-QPC integration are controllable in response to ER occupancy parity. We also show that the 0.7 anomaly, another strongly-correlated state in QPCs, not only has a different physical origin but furthermore counteracts the Kondo spin singlet. These results demonstrate a noninvasive quantum method for sensing spontaneous magnetic impurities within an open nanoconstriction.}


\clearpage

The Kondo effect leads to the screening of a bound electron spin or, equivalently, to fluctuations of the bound spin moment through the formation of a spin singlet with itinerant electron spins. It remains one of the most renowned and challenging problems in condensed matter physics, continually attracting interest and being central to many emergent materials and nanostructures\cite{Im2023,Hartstein2018,Shen2020,Borzenets2020,Madhavan1998,Spinelli2015,Pasupathy2004,Roch2008,Nygard2000,Paaske2006,Goldhaber-Gordon1998,Cronenwett1998,Wiel2000,Noro24,Efferen24}. While Kondo effects have been observed in various mesoscopic systems, their possible presence in quantum point contacts (QPCs)—manifested in Kondo-like zero-bias anomaly (ZBA) peaks in the nonlinear conductance—is particularly intriguing and has created extraordinary interest and debate\cite{Cronenwett2002,Meir2002,Smith2022,Rejec2006,Sfigakis2008,Brun2016,Bauer2013,Schimmel2017,Thomas1996,Kristensen2000,Tokura2002,Rokhinson2006,Yan2018}. Firstly, QPC is a quasi-one-dimensional (1D) constriction where electrons are in principle mobile and hence a localized spin impurity, the prerequisite for the Kondo effect, is not obviously expected. The formation of quasi-bound spin in QPCs was later suggested to be possible through the interplay of strong electron-electron (e-e) interactions and the QPC barrier, which is associated with Friedel oscillations\cite{Rejec2006,Sfigakis2008}. This mechanism is fundamentally different from other Kondo systems such as quantum dots where the spin impurity is unambiguously implanted through electrostatic potential. This thereby makes the Kondo problem in QPC more intricate since the spin impurity itself (if it does appear) is already the consequence of many-body interactions, and how exactly such an interaction-driven quasi-bound spin and its further Kondo-type interaction with surrounding itinerant electrons mutually interact with each other remains elusive. Most studies have so far treated them separately and independently. Secondly, on the other hand, there are also evidences showing that the observed ZBAs are not necessarily the consequence of Kondo mechanism and can also be well interpreted by other mechanisms such as a smeared van Hove singularity at the 1D subband bottom, forming the so-called van Hove ridge in the QPC density of states\cite{Bauer2013,Schimmel2017}. Moreover, the characteristics of conductance anomalies revealed by different studies are diverse, with some supporting the Kondo model and others opposing it\cite{Cronenwett2002,Meir2002,Smith2022,Rejec2006,Sfigakis2008,Brun2016,Bauer2013,Schimmel2017,Thomas1996,Kristensen2000,Tokura2002,Rokhinson2006,Yan2018}, as many-body interactions are sensitive to changes in the QPC potential. The fact that ZBAs in QPCs could be attributed to different scenarios and possess diverse characteristics has made it difficult to identify and reach a conclusion on its real microscopic origin when relying only on inferences from QPC conductance features.

Thirdly, and more importantly, the coexistence of other many-body non-Kondo quantum states in QPCs further adds to the uncertainty, but also presents new opportunities for exploring novel quantum phenomena in these systems. By investigating the interplay between various many-body states in a QPC, one may uncover exciting new possibilities for controlling and manipulating spin dynamics. QPCs have long been a fascinating laboratory for studying many-body physics, and their exceptional controllability of carrier density and interactions has enabled the realization of various spin effects such as Wigner crystallization\cite{Ho2018,Klironomos2007,Meyer2007}, ferromagnetic spin chains\cite{Aryanpour2009}, helical spin transport\cite{Heedt2017}, and spontaneous lifting of spin degeneracy\cite{Rokhinson2006,Chen2008,Chen2012} with coherent and controllable spin manipulation\cite{Chuang2015,Lo2017}. Amongst all the many-body phenomena in QPCs, the most renowned and still debated is a conductance shoulder at approximately $0.7 G_{\text{Q}}$, commonly known as the 0.7 anomaly\cite{Thomas1996}, in addition to the single-particle 1D conductance quantization in steps of $G_{\text{Q}} = 2e^2/h$ (where $e$ is the electron charge and $h$ is Planck's constant). The 0.7 anomaly appears in the linear conductance (that is, at zero source-drain bias) and its presence is usually associated with the aforementioned ZBA peaks in the nonlinear conductance, and hence has been attributed to the same origin\cite{Cronenwett2002,Meir2002,Smith2022,Rejec2006,Sfigakis2008,Brun2016}. However, it has also been suggested that the 0.7 anomaly is a consequence of quasistatic spin texture (spin polarization) instead of dynamic spin fluctuation associated with the Kondo effect\cite{Bauer2013,Schimmel2017,Thomas1996,Kristensen2000,Tokura2002,Rokhinson2006,Yan2018}. This debate has not been resolved since the various proposed interpretations, despite being completely different in their precise nature, all vary with the conductance, density, and interaction strength in a similar way, making it difficult to distinguish between them solely by tuning the parameters of a QPC\cite{Iqbal2013}. It is therefore essential to develop an approach that allows for nonlocal and noninvasive sensing of various QPC quantum states, enabling the examination of their interplay without substantially modifying the QPC parameters.

In this study, we integrate a weakly confined Fabry–P{\'e}rot-type electronic resonator (ER) with controllable quantized electronic states in close proximity to a QPC, and study how the QPC responds to a change in the ER electron occupancy and coupling between them for nonlocal and noninvasive detection of various QPC quantum states. The principle is to have the emergent localized spin in the QPC (if it does exist) coupled with the ER spin state\cite{Rossler2015,Jung23} in our finely set controllable two-impurity Kondo system. This ER-QPC integration, in which the electron number parity and e-e interaction strength are designed to be separately controlled by ER and QPC gates, provides the key to verify and identify different microscopic schemes for QPC conductance anomalies. We distinguish the interaction-driven QPC spin states by their distinct ZBA responses to the ER spins and demonstrate that the nonlocal ER-QPC two-impurity Kondo state collapses when interfering with the 0.7 anomaly. These results shift the current understanding of the origin of conductance anomalies, specifically the 0.7 and ZBA anomalies, from a belief that they stem from the same mechanism, to the one that recognizes their emergence from the competition and coexistence of distinct spin effects.

We employ a weakly confined Fabry–P{\'e}rot (FP)-type ER as an artificial impurity near a QPC, and study how the QPC reacts upon a change in the ER state filling and coupling between them. An off-site ER with odd occupancy contains a bound spin and can have exchange couple with the QPC quasi-bound spin (if it exists) to give rise to the two-impurity (or even-parity) Kondo effect and double-peak ZBA. Removing the net spin moment, as in an ER with even occupancy, the QPC quasi-bound spin alone forms a single-impurity (or odd-parity) Kondo state, resulting in a single-peak ZBA. The ability to control the spin moment using the ER occupancy aids in sensing and understanding the Kondo state intricately dressed by other QPC quantum states.

Device A, comprising a QPC with a spatially separate FP-type ER\cite{Rossler2015,Jung23}, is defined by surface gates above a GaAs/AlGaAs two-dimensional electron gas (Fig.~1a; see Supporting Note 1 for experimental methods). This design facilitates nonlocal and noninvasive quantum sensing of QPC Kondo-related phenomena by independent tuning of the QPC and ER constrictions, which are formed by the gate voltages $V_{\text{qpc}}$ and $V_{\text{f}}$, and $V_{\text{er}}$ and $V_{\text{f}}$, respectively. A Kondo spin within the shallow QPC quasi-bound state (Fig.~1a, bottom inset) and related QPC states, arising from many-body interactions, will all be modulated by the conductance parameter. The FP-type ER constriction (Fig.~1a, dotted arc) reflects itinerant electrons and quantum interference occurs to form discrete energy modes (solid lines) accommodating electron spins (red arrows). Both the QPC and ER constrictions are crucial in establishing the anticipated two-impurity Kondo state, with one impurity located in the QPC and the other in the ER. The typical QPC 1D quantized conductance in steps of $G_{\text{Q}}$ is observed when the ER gate is grounded (Fig.~1b). Figure~1c presents the characteristic measurement results of QPC linear conductance and summarizes the influence of the ER constriction on the QPC behaviour. Adjusting the voltages applied to the QPC and ER forming gates ($V_{\text{qpc}}$ and $V_{\text{er}}$, respectively) produces distinctly different impacts on the QPC conductance. The QPC conductance as a function of $V_{\text{er}}$ shows oscillations superimposed on the quantized conductance traces (red trace), whereas these oscillations are absent or weakened when sweeping $V_{\text{qpc}}$ (purple trace), which distinguishes the impacts of $V_{\text{er}}$ control on ER-QPC constrictions from those due to $V_{\text{qpc}}$ . The observed quasiperiodic conductance oscillations above $G_{\text{Q}}$ are related to the sequential filling of quantized ER modes with $V_{\text{er}}$, which discretely modulates the coupling of QPC to the source reservoir and gives rise to the linear conductance oscillations. In this regime of higher conductance and carrier density, the QPC interaction effects are reduced, enabling a clearer observation of the ER transport properties from the QPC conductance. Conversely, around $G=G_{\text{Q}}$ or below, the conductance oscillations are linked to parity switches of the ER-QPC Kondo state. This is evidenced by the alternating single- and double-peak ZBAs in the nonlinear conductance (inset of Fig.~1c), with further details studied later. We demonstrate that an accidental ER, formed between the neighboring QD and QPC barriers, is more effective at tuning the QPC Kondo spin parity than a distant, strictly confined quantum dot (QD) in the integrated QD-QPC devices T1 and T2 (see Supporting Figs. S1 and S2). The observed parity control of the ER-QPC ZBAs in devices T1 and T2 indicates the formation and function of a working ER and establishes the basis for gate layout and operation in device A with enhanced ER controllability. Moreover, unlike the well-isolated FP-type ER in most previous studies\cite{Rossler2015,Jung23}, the weaker ER constriction in device A allows the ER to be positioned closer to the QPC, enhancing ER-QPC coupling while minimizing its impact on the QPC barrier.

For comparison, in device B, the ER constriction is tuned by finger gates above the split gates, which are electrically isolated by a dielectric layer (Fig.~1d). This dual-gated structure eliminates the spatial separation between the QPC and ER, allowing the ER constriction to invasively interfere with the formation of QPC quasi-bound states and the associated Kondo physics. By varying the ER and QPC gate voltages, the bound state within the QPC can be locally enhanced or suppressed, resulting in a controllable resonance structure in the linear conductance (Fig. 1e). This device design enables tuning of the resonance all the way from $G=0$ towards and then weakening into the $G=G_{\text{Q}}$ quantized plateau.  While the resonance structure may resemble the 0.7 anomaly when above $G=0.5G_{\text{Q}}$, its evolution from $G=0$ suggests a different underlying mechanism. Regarding the nonlinear conductance, only the single-peak ZBA is observed, regardless of the location of the controllable resonance structure (Fig.~1f). We note that, manipulating the interfered ER-QPC constriction using the top finger gates changes the bound-state confinement on-site but does not alter the spatial extent of the QPC potential hill. As a result, only the ZBA peak height is modulated, and a two-impurity Kondo state and ZBA peak splitting do not occur in device B.

Many different approaches have been proposed to detect a quasi-bound spin in a QPC, including the use of multiple pairs of split gates and scanning gate microscopy\cite{Iqbal2013,Brun4290,Yoon2009,Yoon2012,Fransson2014}. However, the details remain unclear, as these methods detect only the charge component of the quasi-bound state or significantly distort the QPC potential (see further discussion in Supporting Note 2). Building on insights into two-impurity QD Kondo systems\cite{Jeong2221,Craig2004,Chen2004,Simon2005,Potok2007,Keller2015,Bork2011,Pruser2014,Esat2016,Duan2020,Yeh2020} and our proposed gate layout, we are able to examine how the QPC quantum state reacts to the spatially separate ER spins (Fig.~1a) by tracing the ZBAs with systematic $V_{\text{qpc}}$ and $V_{\text{er}}$ control over the QPC and ER constrictions in device~A. Figures~2a and 2b show the linear conductance as a function of $V_{\text{qpc}}$ at various $V_{\text{er}}$ settings to change the ER electron occupancy. The first quantized plateau oscillates in $G$ with $V_{\text{er}}$ and this oscillation is reminiscent of the alternation between single- and two-impurity Kondo effect, which respectively increases and decreases the linear conductance, similar to the cases in double QDs\cite{Jeong2221,Craig2004}. The oscillations diminish as a more negative bias depletes the ER states and reduces the ER-QPC coupling (Fig.~2b). The odd-even conductance modulation under the optimized $V_{\text{f}}$ setting in Fig.~2a is emphasized by plotting transconductance $\partial G/\partial V_{\text{qpc}}$ as a function of $G$ and $V_{\text{er}}$ (Fig.~2c). Regions shown in dark blue (low transconductance) correspond to where the plateau exists. Note that there is an additional plateau-like feature below the first plateau, which is identified as the 0.7 anomaly; however, it does not oscillate but monotonically decreases to $0.5G_{\text{Q}}$ with $V_{\text{er}}$ (along the arrow direction), in stark contrast to that of the first plateau. The fact that the 0.7-anomaly conductance does not oscillate in the same manner as the alternating single- and double-peak ZBAs with switching the ER occupancy parity implies the presence of an additional mechanism—referred to as 0.7 anomaly physics—that influences QPC Kondo-related behavior. To further explore this, we investigate the nonlinear conductance. Figure~2d shows that single- and double-peak ZBAs appear alternately with successively changing the ER occupancy (equivalent to turning on and off a net ER spin moment) when the conductance is close to $G_{\text{Q}}$ (bold traces). The switch cycle for linear and nonlinear conductance features are in phase, i.e., the presence of single- and double-peak ZBAs in the nonlinear conductance coincide with the maximum and minimum extremes of oscillations in linear conductance of the first plateau (denoted in solid and open symbols for single- and two-impurity Kondo state, respectively, in Figs.~2a, 2c, and 2d). This result is significant in that it provides conclusive evidence of the existence of QPC quasi-bound spin which is nonlocally accessible. We also notice that when $G$ is lowered to the 0.7-anomaly region (dashed traces in Fig.~2d), the ZBAs weaken and are all in the form of single peak no matter whether the ER has a zero or non-zero spin moment. This suggests a counteracting relationship between the coexisting Kondo effect and 0.7-anomaly physics.

Figure~3a presents the interplay between the 0.7 anomaly and Kondo ZBAs by showing $G$ as a function of $V_{\text{er}}$ at various $V_{\text{qpc}}$ settings. The QPC conductance oscillates quasiperiodically with $V_{\text{er}}$ due to ER state filling\cite{Jung23}, and these oscillations can penetrate into the 0.7-anomaly region. The corresponding evolution of nonlinear conductance is also crucial. For clarity, we also present the second derivative of the nonlinear conductance $-\partial^{2} G/\partial V_{\text{sd}}^{2}$ (Figs.~3b-3d) for the analyses, which removes the rising background of 1D conductance and provides a better method to locate and investigate the ZBA peak structures (see Supporting Fig.~S3a).

The resonant line structure observed in the $-\partial^{2} G/\partial V_{\text{sd}}^{2}(V_{\text{sd}}, V_{\text{er}})$ map at $V_{\text{qpc}}=-1.60$~V for $V_{\text{er}} > -1.4$~V provides compelling evidence for the control of interference-induced quantized states in the FP-type ER (Fig.~3b). The obtained energy level spacing  $\delta_{\text{er}}\approx 0.3$~meV gives an estimation of the ER size $L\approx 200$~nm by $\delta_{\text{er}}\sim\hbar^{2}\pi^{2}/m^{\text{*}}L^{2}$ (where $m^{\text{*}}=0.067m_{\text{e}}$ is the electron effective mass)\cite{Jung23,Rossler2015} within its lithographic scale of $L_{\text{er}}\approx400$~nm (Fig.~1a). In this region, the oscillatory QPC conductance indicates the filling of ER states (upper green bold trace in Fig.~3a). Although tuning $V_{\text{er}}$ changes the ER occupancy by two electrons (i.e., a spin-zero charge state), odd occupancy (i.e., a non-zero spin state) can emerge due to coupling with a nearby QPC bound spin (if it exists), forming an anticipated two-impurity Kondo state and double-peak ZBA, as demonstrated in the reported ER-QD system\cite{Rossler2015}. In the ER-QD system, the QD spin impurity is unambiguously implanted, allowing precise control over the ER-QD Kondo state. In contrast, Kondo phenomena in the ER-QPC system are more complex because the QPC spin impurity itself arises as a consequence of many-body interactions and is highly sensitive to QPC parameter tuning. When entering the tunneling regime below the first plateau at $V_{\text{er}}=-1.4$~V, the conductance oscillations become complex due to enhanced many-body interactions. Interestingly, switching between the single- and double-peak ZBAs in response to ER occupancy changes only occurs when the conductance is sufficiently high or low, i.e., near the first plateau or below $0.5G_{\text{Q}}$ (top and bottom insets in Fig.~3b, respectively). The ZBA retains single-peak character around $V_{\text{er}}=-1.8$~V (middle inset in Fig.~3b), the regime that interferes with the 0.7 anomaly (shaded area in Fig.~3a). We continue to probe the observed counteracting relationship between 0.7-anomaly physics and Kondo effect by investigating the linear and nonlinear conductance against $V_{\text{er}}$ at $V_{\text{qpc}}=-1.77$~V. The linear conductance at $V_{\text{qpc}}=-1.77$~V is characterized in Fig.~3a (lower green bold trace), where there are more oscillation cycles due to manipulation of ER occupancy within the 0.7-anomaly conductance region (shaded area). Figure~3c shows the $-\partial^{2} G/\partial V_{\text{sd}}^{2}(V_{\text{sd}}, V_{\text{er}})$ mapping data while Fig.~3d shows the line-cut traces at increasing values of $V_{\text{er}}$, indicated by circles in Fig.~3c, along the arrow direction. The successive parity switches of ER occupancy cause the ZBA to alternate between single and double peaks (red solid and blue open circles, respectively, in Fig.~3c, and corresponding red and blue traces in Fig.~3d) when $G < 0.5G_{\text{Q}}$ (i.e., $V_{\text{er}}<-1.1$~V). However, this alternation disappears when entering the region of 0.7 anomaly ($V_{\text{er}}>-1.1$~V). The ZBAs in this region remain as a single peak regardless of the ER spin moment and only their peak height (vertical bars) is modulated in an alternating fashion\cite{Potok2007}. The controllable ER occupancy offers a means for noninvasive sensing of QPC quantum states. Our results show that the quasi-bound spin is no longer accessible to its nearby ER spin and/or itinerant electrons to form the Kondo effect when the 0.7-anomaly physics coexists in the QPC. In contrast to the argument that the 0.7 conductance anomaly arises from the Kondo effect\cite{Meir2002,Cronenwett2002,Rejec2006}, we show the complete opposite that the 0.7-anomaly physics actually hampers QPC Kondo spin fluctuations, resulting into a suppressed ZBA peak height and splitting. These findings support that the observed 0.7 conductance anomaly arises from the interplay of different coexisting QPC quantum states. They also align with the prediction that the 0.7 anomaly occurs at a conductance and carrier density regime where e-e interactions are strongest, promoting QPC spin polarization\cite{Jaksch2006,Lassl2007} and consequently weakening ER-QPC singlet coupling.

Finally, we demonstrate the temperature and magnetic field dependence of the single- and double-peak ZBAs in device~A. As the temperature increases, the double-peak ZBA (Fig.~4a) is smeared into a broad peak at $T=0.6$~K, while the single-peak ZBA (Fig.~4b) is still present (dashed traces). The single-peak ZBA eventually disappears at $T=1.1$~K (bold traces), which is significantly lower than the approximately 10 K predicted by the Kondo model\cite{Cronenwett2002}. The Kondo correlation can also be examined by an in-plane magnetic field $B$. Figure~4c shows the re-emergence of single-peak ZBA from the double-peak ZBA at $B=3$~T (dashed trace). As $B$ increases further, only peak suppression is observed without further splitting for both ZBA parities (Figs.~4c and 4d). The absence of double-peak ZBAs may result from the thermal broadening of ER states ($4k_{\text{B}}T\approx0.2$~meV at $T=0.6$~K, where $k_{\text{B}}$ is the Boltzmann constant) and Zeeman effect (whose contribution cannot be estimated in this work due to the undetermined Land{\'e} $g$-factor in the presence of strong interactions\cite{Vionnet2016}). The detailed peak evolution with either temperature and magnetic field depends on the delicate competition between single- and two-impurity Kondo state, which is further complicated by strong interactions in QPCs (see Supporting Note 4 for further discussion).

The microscopic origin of the 0.7 anomaly together with the possible existence of Kondo effect in QPC is one of the most intriguing and challenging problems in mesoscopic physics, and has been debated for more than two decades. Our tunable ER-QPC designs and experiments bring fresh insights into this fundamental problem. We directly demonstrate the existence of a quasi-bound spin and Kondo correlation in the QPC by alternately forming the single- and two-impurity Kondo effects via nonlocal ER spin moment control. However, in stark contrast to the proposal that if the Kondo spin fluctuations exist in QPCs they are responsible for causing the 0.7 anomaly, we show that they are two separate entities and can coexist with each other. Our nonlocal and noninvasive access to the QPC quantum state is essential to exploring such a quantum system, which has more than one many-body phenomena coexisting within it. It allows us to manipulate and track one particular effect without significantly disturbing the other, by which we can separate the different mechanisms and explore the interplay among them. It reveals that the 0.7 quantum state behaves as a quasistatic spin texture and counteracts the fast Kondo spin fluctuation. Our results provide a versatile route for not merely investigating the interactions between localized and itinerant spins in open nanoconstrictions, but also for control and detection of elusive spin states in QPCs or other quantum systems, which may open up new possibilities in semiconductor spintronics and quantum engineering.

\section*{ASSOCIATED CONTENT}
\textbf{Supporting Information}
\\Experimental methods, and extended data and analyses for devices~A, T1, and T2.

\section*{AUTHOR INFORMATION}
\textbf{Author Contributions}
\\S.-T.L., C.-C.W., J.-H.C., and M.-W.C. performed the measurements and analysed the data with help from S.-C.H., S.-H.C., Y.-C.H., P.-T.W, Y.-C.W., and C.-T.L. G.L.C., S.-T.L., S.-C.H., and L.W.S. fabricated the devices with contributions from M.P., and T.-M.C; I.F. and D.A.R. provided wafers; J.P.G. and G.A.C.J. performed electron beam lithography. S.-T.L. and T.-M.C. designed and coordinated the study, and wrote the paper with input from all authors.

\noindent \textbf{Notes}
\\The authors declare no competing financial interest.

\section*{ACKNOWLEDGMENTS}
We thank Hsiang-Feng Sun and Yu-Chih Su for helpful discussions. This work was supported by the National Science and Technology Council (Taiwan), the Headquarters of University Advancement at the National Cheng Kung University, and the Engineering and Physical Sciences Research Council (UK). S.-T.L. acknowledges additional support from the Higher Education Sprout Project of the National Yang Ming Chiao Tung University and Ministry of Education (Taiwan).

\newpage

\linespread{1.2} \selectfont
\begin{figure}
\begin{center}
\includegraphics[width=1\columnwidth]{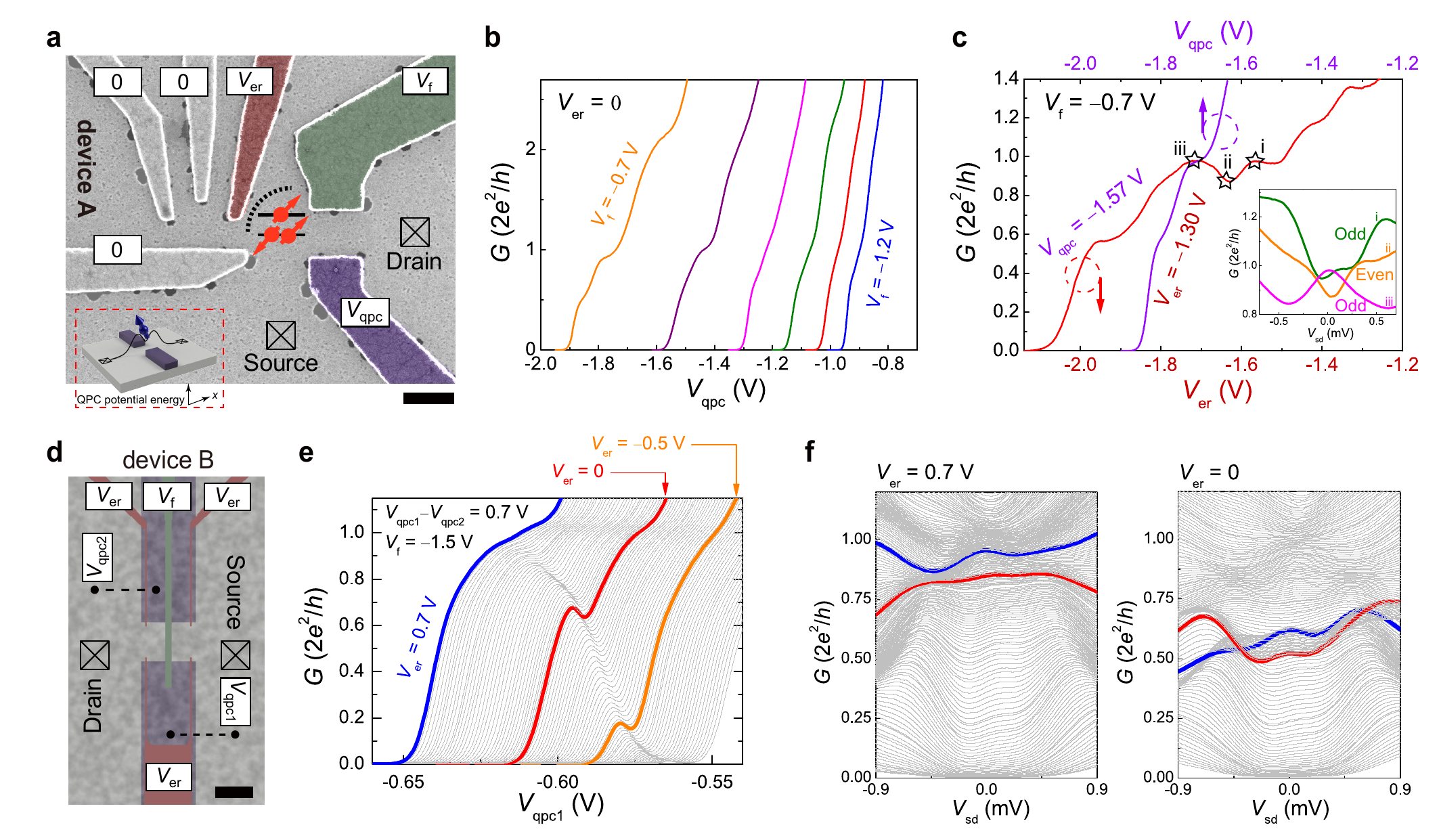}
\end{center}
\caption{\small \textbf{Integrated electronic resonator-quantum point contact device and its linear and nonlinear transfer characteristics.} (\textbf{a}) False-color scanning electron
micrograph of the gate pattern for device~A. The quantum point contact (QPC) and electronic resonator (ER) are primarily controlled by voltages $V_{\text{qpc}}$, $V_{\text{er}}$, and $V_{\text{f}}$ with all other gates grounded. Bottom inset, a schematic representation of the QPC potential profile and an interaction-driven quasi-bound spin. Middle inset, a schematic representation of ER spin states. (\textbf{b}) Linear conductance $G$ as a function of $V_{\text{qpc}}$ with $V_{\text{er}}=0$. (\textbf{c}) Linear $G$ as a function of $V_{\text{er}}$ ($V_{\text{qpc}}$) at the selected $V_{\text{qpc}}$ ($V_{\text{er}}$). Inset, nonlinear $G$ at three different $V_{\text{er}}$ values, marked by stars. (\textbf{d}) Lithographic gate pattern of device~B, illustrating the dual-gated structure. The two gate layers are separated by a SiO$_{\text{2}}$ dielectric. The gates are biased with the indicated voltages. (\textbf{e}) Linear $G$ against $V_{\text{qpc1}}$ (while co-sweeping $V_{\text{qpc2}}$ at a fixed $V_{\text{qpc1}}-V_{\text{qpc2}}$) with changing $V_{\text{er}}$. (\textbf{f}) Nonlinear $G$ against $V_{\text{qpc1}}$ at two selected $V_{\text{er}}$ values, with the same $V_{\text{qpc1}}-V_{\text{qpc2}}$ as in \textbf{e}. The scale bars in \textbf{a} and \textbf{d} correspond to 400 nm.}
\end{figure}

\clearpage
\begin{figure}
\begin{center}
\includegraphics[width=1\columnwidth]{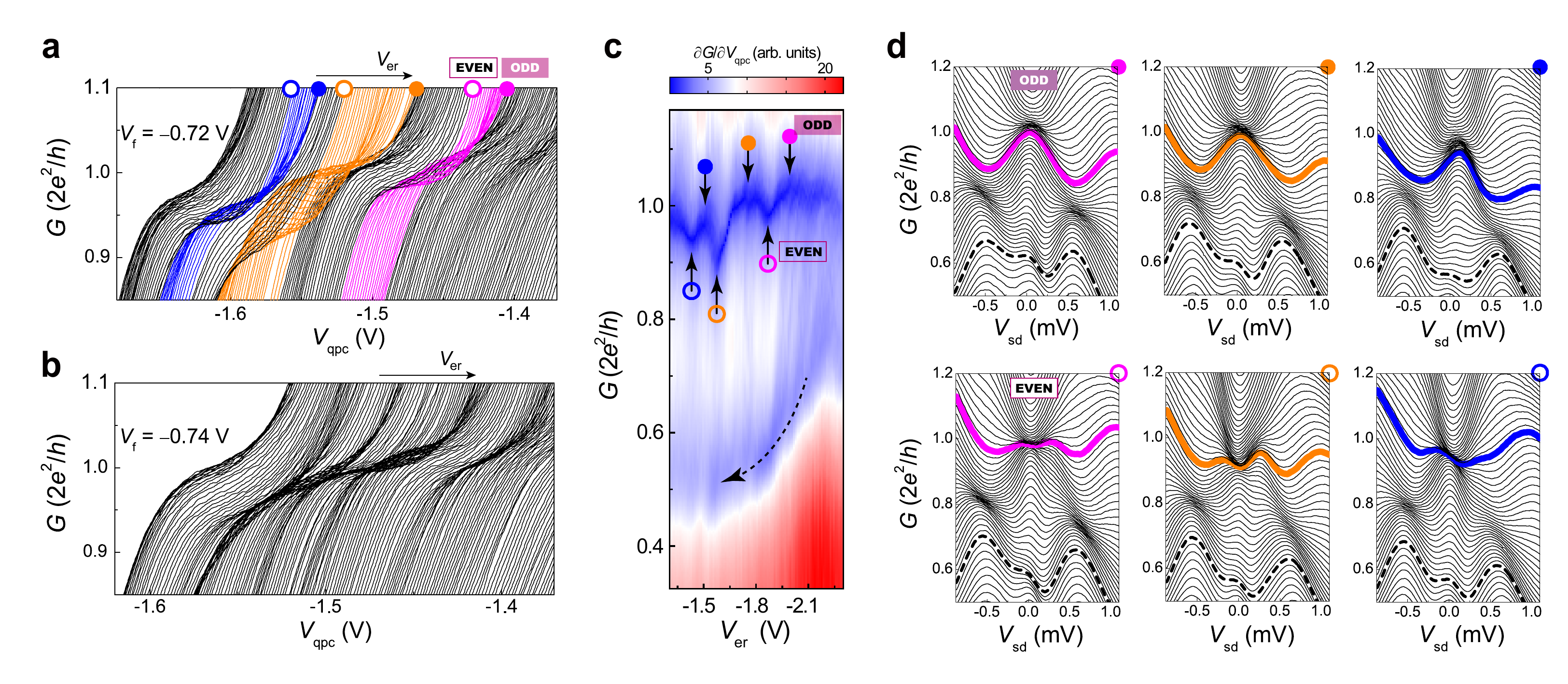}
\end{center}
\caption{\small \textbf{Transport anomalies in linear and nonlinear conductance, device A.} (\textbf{a, b}) Sequences of linear conductance traces $G(V_{\text{sd}}=0, V_{\text{qpc}})$ with decreasing $V_{\text{er}}$ along the arrow direction for $V_{\text{f}}=-0.72$~V (\textbf{a}) and $-0.74$~V (\textbf{b}). (\textbf{c}) Transconductance $\partial G/\partial V_{\text{qpc}}$ versus $G$ and $V_{\text{er}}$, displaying the oscillating and monotonic dependence of the first plateau and 0.7 anomaly on $V_{\text{er}}$, respectively. (\textbf{d}) Nonlinear conductance traces $G(V_{\text{sd}}, V_{\text{qpc}})$ at various $V_{\text{er}}$ settings. Six circles marked in (\textbf{a}), (\textbf{c}), and (\textbf{d}) indicate the data at six characteristic gate voltages $V_{\text{er}}$, where the first plateau shows local extremes (solid and open circles for maxima and minima, respectively). These are accompanied by single- and double-peak ZBA in the nonlinear conductance, respectively [bold traces in (\textbf{d})]. Dashed traces in (\textbf{d}) highlight the suppressed single-peak ZBA around the 0.7 anomaly.}
\end{figure}

\begin{figure}
\begin{center}
\includegraphics[width=1\columnwidth]{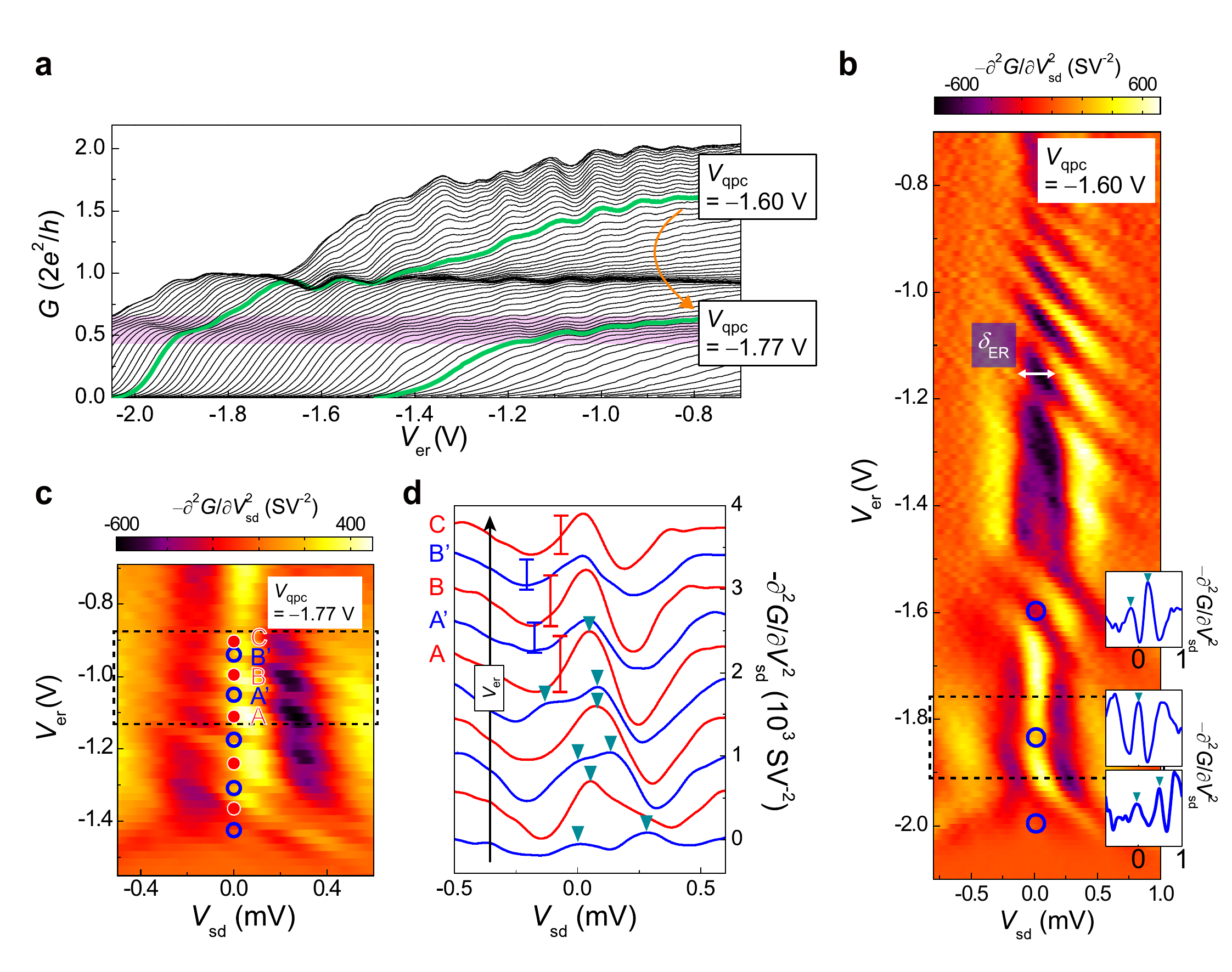}
\end{center}
\caption{\small \textbf{Interplay of the QPC Kondo and 0.7 anomaly in device~A.} (\textbf{a}) Linear conductance versus $V_{\text{er}}$ as $V_{\text{qpc}}$ is decreased along the arrow direction with $V_{\text{f}}$ fixed at $-0.7$~V. Green traces correspond to $V_{\text{qpc}}=-1.60$~V and $-1.77$~V, respectively. The plateaus (at $G_{\text{Q}}$ and $0.5G_{\text{Q}}$) appear as darker regions with a higher density of traces. The shaded areas indicate the simultaneous appearance of the conductance oscillations and 0.7 anomaly. (\textbf{b}) Color map of the negative second derivative $-\partial^{2} G/\partial V_{\text{sd}}^{2}$ versus $V_{\text{er}}$ and $V_{\text{sd}}$ at $V_{\text{qpc}}=-1.60$~V. Insets, line cuts of the data at three selected $V_{\text{er}}$, which give the local minima of their zero-bias values (open circles). Dashed rectangle highlights the region of the 0.7 anomaly. (\textbf{c}) Color map of $-\partial^{2} G/\partial V_{\text{sd}}^{2}$ versus $V_{\text{er}}$ and $V_{\text{sd}}$ at $V_{\text{qpc}}=-1.77$~V. The value at $V_{\text{sd}} = 0$ oscillates as a function of $V_{\text{er}}$. (\textbf{d}) Line cuts of the data in (\textbf{c}) at increasing $V_{\text{er}}$ along the arrow direction [marked with solid and open circles in (\textbf{c}) for different ER occupancy parities]. Traces alternate between single and double ZBA peaks in $-\partial^{2} G/\partial V_{\text{sd}}^{2}$ (triangles) while they exhibit only single-peak character with peak height modulation in the 0.7-anomaly region [dashed rectangle in (\textbf{c}) and the corresponding top five traces in (\textbf{d})]. Traces are offset from bottom for clarity. Vertical bars are shown to compare the zero-bias peak heights at different $V_{\text{er}}$.}
\end{figure}

\begin{figure}
\begin{center}
\includegraphics[width=0.7\columnwidth]{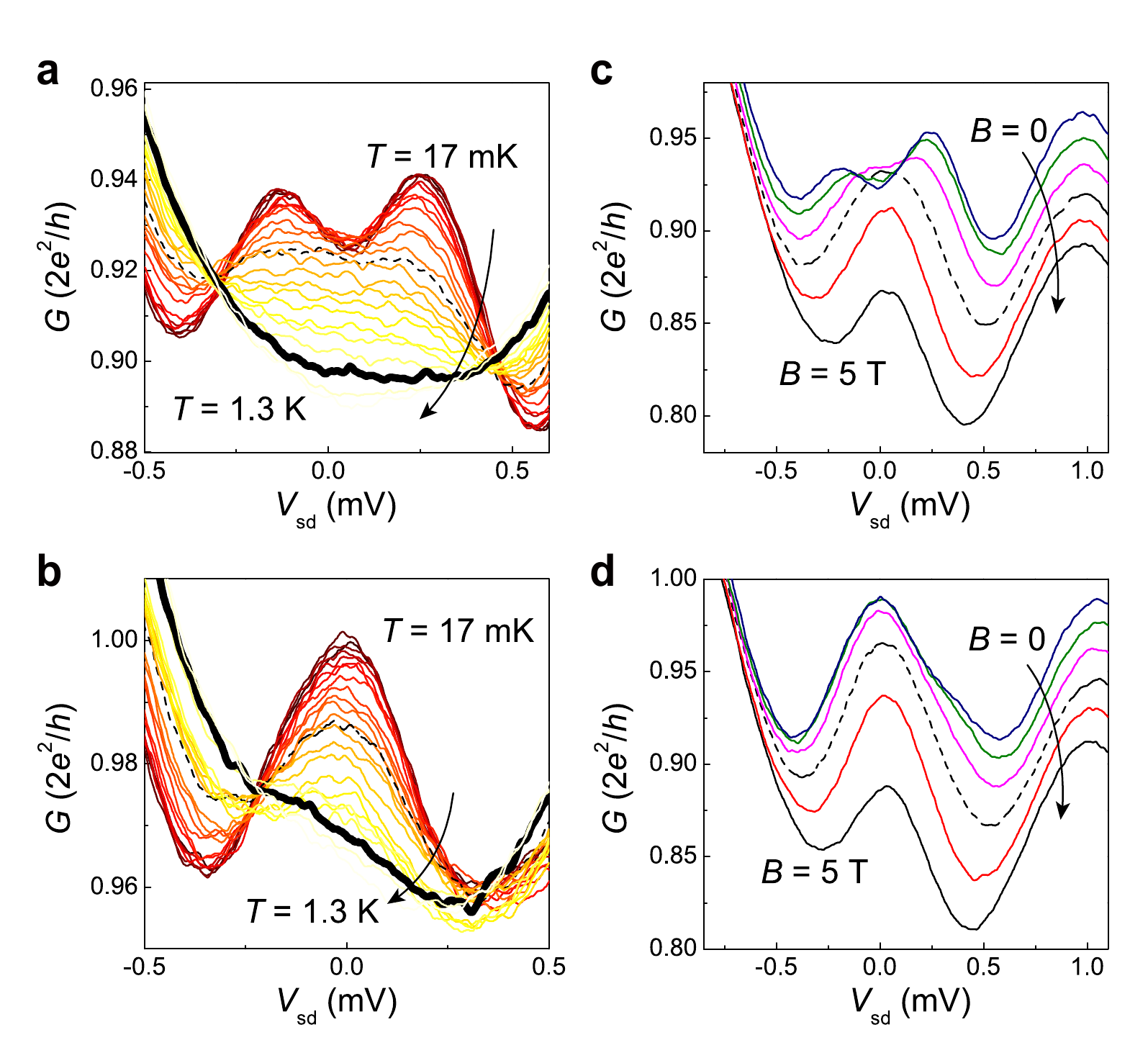}
\end{center}
\caption{\small \textbf{Destruction of Kondo spin singlet in device~A.} (\textbf{a, b}) Double- and single-peak ZBAs with increasing temperature $T$ along the arrow direction. The ZBA peaks are completely suppressed when $T = 1.1$~K. Dashed and bold traces indicate the data at $T = 0.6$~K and $1.1$~K, respectively. (\textbf{c, d}) Double- and single-peak ZBAs at various in-plane magnetic fields from $B=0$ to $5$~T in steps of $1$~T. The dashed traces indicate the data at $B = 3$~T.}
\end{figure}

\end{document}